\documentclass[prb,twocolumn,amsmath,amssymb]{revtex4}

\begin{document}


\title{Superconducting single-electron transistor and the 
$\phi$-modulation of supercurrent}


\author{M. Aunola}
\affiliation{Department of Physics, University of Jyv\"askyl\"a,
P.O. Box 35 (YFL), FIN-40351 Jyv\"askyl\"a, FINLAND}


\date{\today}

\begin{abstract}
An analytical expression for the supercurrent of a superconducting 
single-electron transistor (SSET) is derived. The derivation is
based on analogy between the model Hamiltonian 
for $E_{\mathrm{J}}> E_{\mathrm{C}}$ and a
discrete, one-dimensional harmonic oscillator (1DDHO). 
The resulting supercurrent is nearly identical 
to the supercurrent obtained from a continuous harmonic
oscillator Hamiltonian.

\end{abstract}


\maketitle


The superconducting single-electron transistor consists of two
consequent Josephson junctions and an intervening island on
which the amount of charge can be controlled by a gate voltage.
The relevant energy scales of the system are given by the Josephson
energy $E_{\mathrm{J}}$ and the
charging energy $E_{\mathrm{C}}:=(-2e)^2/[2(C_1+C_2+
C_{\mathrm{g}})]$ where
$C_1$, $C_2$ and $C_{\mathrm{g}}$ are the capacitances 
of the two junctions and the gate capacitance, respectively.
For independent junctions 
the Hamiltonian of the system is given by
\vskip-0.6truecm
\begin{equation}
  H=H_{\mathrm{C}}-\sum_{j=1}^2E_{\mathrm{J,j}}\cos(\phi_j),
\end{equation}
\vskip-0.2truecm
\noindent{where} 
$H_{\mathrm{C}}$ gives the charging energy of the island
and $\phi_j$ is the phase difference across the $j^{\mathrm{th}}$
junction. 


The proper variables for description of the system are the phase
difference across the array $\phi=\phi_1+\phi_2$, and the number of
Cooper pairs on the island $N$.\cite{ave91} The phase
difference $\phi$
is a constant of motion if the voltage across the SSET is
ideally biased to zero. The
Hamiltonian is fixed using the arguments given in 
Ref.~\onlinecite{aun00}, i.e. by 
taking $C_j=c_jC$ and $E_{\mathrm{J},j}=
c_jE_{\mathrm{J}}$, where $c_1^{-1}+c_2^{-1}=2$. 
The normalised gate charge $q=V_{\mathrm{g}}C_{\mathrm{g}}/(-2e)$ 
sets the amount of free charge on the island to $(-2e)(N-q)$.
In the charge state representation the Hamiltonian reads\cite{eil94,tink96}
\vskip-0.5truecm
\begin{eqnarray}
H_{\mathrm{C}}&=&E_{\mathrm{C}}\sum_N(N-q)^2\vert N,\phi\rangle
\langle N,\phi\vert,\\
H_{\mathrm{J}}&=&
-(E_{\mathrm{J}}/2)(c_1^2+c_2^2+2c_1c_2\cos(\phi))^{1/2}\cr
&&\ \ \times\sum_N\left(e^{i\theta(\phi)}\vert N+1,\phi\rangle
\langle N,\phi\vert+\,\mathrm{h.c.}\right),
\end{eqnarray} 
\vskip-0.1truecm
\noindent{where} $\tan(\theta)=(c_1-c_2)\tan(\phi/2)/(c_1+c_2)$. In
 the limit of vanishing charging energy when
$H_{\mathrm{C}}\rightarrow 0$, the ground state energy and 
supercurrent are given by\cite{tink96}
\vskip-0.5truecm
\begin{eqnarray}
E(\phi)\hskip-5pt&=&\hskip-3pt
-E_{\mathrm{J}}(c_1^2+c_2^2+2c_1c_2\cos(\phi))^{1/2},\\ 
I^{(0)}_{\mathrm{S}}&=&\hskip-3pt
\frac{-2e}\hbar\frac{\partial E(\phi)}
{\partial\phi}=\hskip-1pt
\frac{(-2e/\hbar)E_{\mathrm{J}}c_1c_2\sin(\phi)}
{(c_1^2+c_2^2+2c_1c_2\cos(\phi))^{1/2}}.\label{eq:super}
\end{eqnarray}
\noindent{If} the charging effects are not negligible the Hamiltonian
$H_{\mathrm{C}}+H_{\mathrm{J}}$ expressed in unit of $E_{\mathrm{C}}$
is identical to that of a 1DDHO with coupling constant 
$\varepsilon_{\phi}:=E(\phi)/E_{\mathrm{C}}$. The
eigenenergies are independent of the phase factor $e^{i\theta(\phi)}$
which simply fixes the relative
phase between consequtive charge states $\vert N\rangle$ and
$\vert N+1\rangle$.

For a continuous HO with the same  $\varepsilon_{\phi}$
the eigenenergies are given by $E_j=-\varepsilon_{\phi}+
\sqrt{2\varepsilon_{\phi}}(j+\frac12)$.
In case of a 1DDHO with large $\varepsilon_{\phi}$
the bottom of the well is lifted by 
approximately $\frac18$ and oscillator frequency $\sqrt{2a}$ is 
replaced by $\sqrt{2a}-\frac18$. With these modifications numerically
obtained eigenstates satisfy the virial theorem $\langle H_{\mathrm{J}}
\rangle=\langle H_{\mathrm{C}}\rangle$ quite well. The agreement
is best for the ground state for which the expression
\vskip-0.7truecm
\begin{equation}
E_{0}({\varepsilon_{\phi}})
=-\varepsilon_{\phi}+\sqrt{\varepsilon_{\phi}/2}+1/16\label{eq:e0}
\end{equation}
\vskip-0.2truecm
\noindent{very} accurate for $\varepsilon_{\phi}>10$ and even at $
\varepsilon_{\phi}\approx2$ the error is smaller than 
$0.01$ for any $q$. Because of the constant correction
the derivative $\partial E_0/\partial \varepsilon_{\phi}$ 
is the same as in the continuous case.
For weaker couplings with $\varepsilon_{\phi}\lesssim2$ the
minimum position $q$ of the potential energy becomes important,
but direct diagonalisation of the Hamiltonian is simple.
When Eq.~(\ref{eq:e0}) is valid we obtain the final result 
\vskip-0.6truecm
\begin{equation}
I^{\mathrm{SSET}}_{\mathrm{S}}(\phi)=
I^{(0)}_{\mathrm{S}}[1-(8\varepsilon_{\phi})^{-1/2}],
\label{eq:super2}
\end{equation}
\vskip-0.2truecm
\noindent{where} $I^{(0)}_{\mathrm{S}}$ is the supercurrent in the
absence of charging effects. The magnitude of the correction 
$(8\varepsilon_\phi)^{-1/2}$ is of the order of $10$ \%  when
$\varepsilon_\phi\sim10$.
The correction slightly decreases the maximal obtainable
supercurrent and it is important for  nearly homogeneous arrays 
($c_1\approx 1$) as the coupling strength $E_\phi$ becomes 
small near $\phi=(2k+1)\pi$, where $k$ is an integer.

This work has been supported by the Academy of Finland
under the Finnish Centre of Excellence Programme 2000-2005
(Project No. 44875, Nuclear and Condensed Matter Programme at JYFL).
The author thanks Dr. S.~Paraoanu for insightful comments.

\bibliography{SSET}

\end{document}